\documentclass[conference]{IEEEtran}
\usepackage{amsmath,amsthm,amssymb,graphicx,epstopdf}

\newtheorem{theorem}{Theorem}[section]
\newtheorem{corollary}{Corollary}[section]
\newtheorem{lemma}{Lemma}[section]

\def \St{{{\cal S}^2}}
\def \bS{{\beta_\So}}

\def \om{{\omega}}
\def \omh{{\hat{\omega}}}

\def \So{{S_\omega}}

\def \St{{\tilde{S}}}
\def \Sp{{S^\prime}}

\def \SNR{{\text{SNR}}}
\def \Ist{{I_\St}}
\def \Isto{{\bar{I}_\St}}

\def \IsttS{{I(X^{(t)}_{S \setminus \St} ; Y^{(t)} | X^{(t)}_\St, \beta_S)}}

\title{Information-Theoretic Bounds for \\ Adaptive Sparse Recovery }

\author{\IEEEauthorblockN{Cem Aksoylar} \IEEEauthorblockA{Department of Electrical and \\ Computer Engineering, \\ Boston University, Boston, MA 02215} \and \IEEEauthorblockN{Venkatesh Saligrama} \IEEEauthorblockA{Department of Electrical and \\ Computer Engineering, \\ Boston University, Boston, MA 02215}}

\begin{document}

\maketitle

\begin{abstract}
  We derive an information-theoretic lower bound for sample complexity in sparse recovery problems where inputs can be chosen sequentially and adaptively. This lower bound is in terms of a simple mutual information expression and unifies many different linear and nonlinear observation models. Using this formula we derive bounds for adaptive compressive sensing (CS), group testing and 1-bit CS problems. We show that adaptivity cannot decrease sample complexity in group testing, 1-bit CS and CS with linear sparsity. In contrast, we show there might be mild performance gains for CS in the sublinear regime. Our unified analysis also allows characterization of gains due to adaptivity from a wider perspective on sparse problems. 
  
\end{abstract}

\section{Introduction}

For many sparse recovery applications, it has been shown that adaptive methods with sequential and flexible measurement designs improve practical performance compared to nonadaptive methods. From a theoretical point of view, adaptive methods should perform at least as well as nonadaptive methods asymptotically, as the latter is a special case of the former. However, it is an interesting problem to determine whether they can perform strictly better for different recovery problems and problem conditions. While such methods have been theoretically analyzed for specific problems of interest, it is not clear at a high level what properties of sparse problems allow adaptive methods to have strictly better recovery performance compared to nonadaptive ones.

In this work we consider a high-level unifying framework and obtain a lower bound on the sample complexity of adaptive sparse recovery problems. Our framework characterizes the problems of interest with the following sparsity assumption: 
Let $X = (X_1, \ldots, X_N)$ be a set of variables and $Y$ be a corresponding observation generated by an observation model $P(Y | X)$ which satisfies the conditional independence property
\begin{equation}\label{eq:markov}
  P(Y | X) = P(Y | \{X_k\}_{k \in S}), 
\end{equation}
for some set of variables $S \subset \{1,\ldots,N\}$, where $|S| = K$. Then, the aim is to determine the salient set $S$ given $T$ samples of variable/observation pairs, $(X^T, Y^T)$. These pairs are generated sequentially and variables $X$ can be chosen \emph{adaptively} depending on past variables and observations. This framework encapsulates many sparse problems of interest, e.g.\ support recovery in compressive sensing (CS) \cite{donoho}, its nonlinear extensions (e.g.\ quantized CS \cite{1bit}) and other nonlinear problems such as group testing \cite{group_testing}.
\begin{figure}[h]
  \centering
  \includegraphics[width=0.40\textwidth,clip=true,trim=2cm 8cm 2.5cm 6cm]{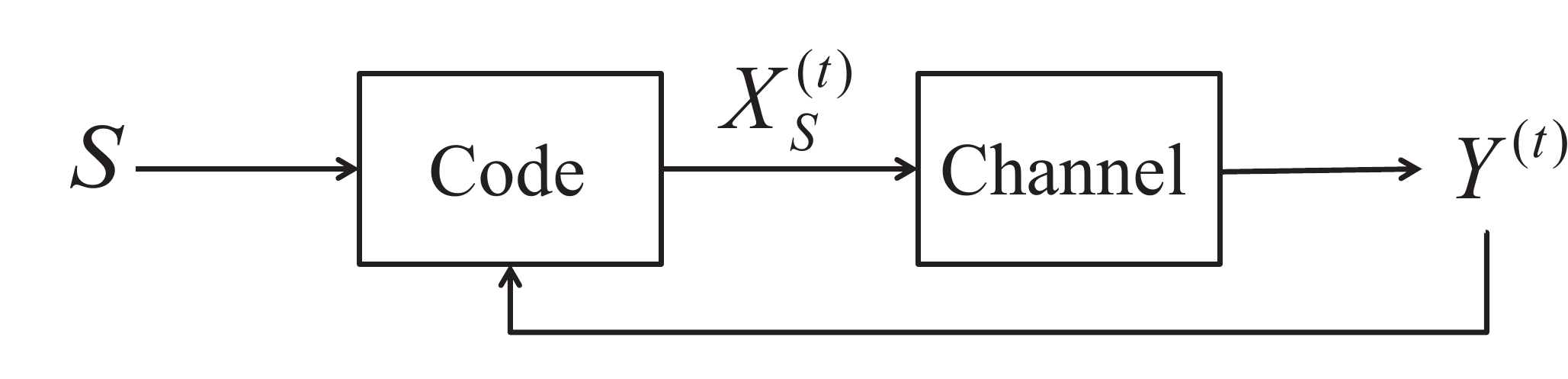}
  \caption{Channel model representation of adaptive sparse recovery.} 
\end{figure}

As the result of our analysis, we obtain a mutual information formula for the lower bound, which depends on the observation model $P(Y | X_S)$ and the distribution of $X$, $p_t(X)$, at each step $t$ of the sequence. We obtain this bound using a Fano's inequality type argument, inspired by the proof of the upper bound on capacity for channel coding with feedback \cite{coverbook}. Our result is unifying for all adaptive sparse recovery problems, similar to  \cite{arxiv,arxiv_dep,ssp,strong}. These works consider the nonadaptive case where variables are generated by a distribution $p(X)$ IID over $T$ samples.

We then obtain adaptive lower bounds for linear and nonlinear applications. We look at the highly nonlinear problem of group testing, where we show that $T = \Omega(K \log(N/K))$ tests are necessary for the adaptive case. This bound can be achieved by nonadaptive methods \cite{group_testing}, implying adaptivity cannot improve performance. Similarly, we consider 1-bit CS as a nonlinear extension of linear CS. We again show $T = \Omega(K \log(N/K))$ is a lower bound to argue adaptivity cannot help, as it is matched by nonadaptive upper bounds \cite{gupta,arxiv} for sufficiently high SNR. 
The same phenomenon happens for linear CS with linear sparsity $K = \Theta(N)$. In this case we show $\SNR = \Omega(\log N)$ is necessary and the lower bound $T = \Omega(N)$ is achieved by nonadaptive methods for this SNR \cite{shuchin}. However for sublinear sparsity, we show there might be mild gains for $T$ and SNR, consistent with the results of \cite{malloy,haupt}. 

There is a large body of work on both adaptive recovery methods and lower bounds, however these analyses are fragmented compared to our unifying approach as they only consider specific problems. The linear problem of adaptive CS has been especially well-studied: Lower bounds have been derived for support recovery \cite{davenport,castro} and adaptive algorithms have been analyzed to obtain upper bounds \cite{malloy,haupt}. There is relatively little work on adaptive recovery on nonlinear models. Adaptive group testing has been investigated by \cite{aldridge,aldridge2} where lower bounds are derived and adaptive algorithms are analyzed (see ref.s in \cite{aldridge2}). Adaptive 1-bit CS algorithms have been proposed \cite{1bit}, however adaptive lower bounds have not been studied to the extent of our knowledge. 

The generality of our analysis also allows us to look at the big picture and comment on the gains due to adaptivity and how it is related to the nature of a problem. We conjecture that adaptivity may help only if there are ``sum-power''-like constraints on the variables as in linear CS. If the variables are not constrained and the difficulty of the problem only stems from the observation model, adaptivity does not increase asymptotic performance.

\section{An Information-Theoretic Framework for Sparse Recovery}

We assume that an observation $Y$ is generated by an observation model for which we assume $P(Y | X) = P(Y | X_S)$, for $S \subset \{1,\ldots,N\}$ with $|S| = K$ and $X_S = \{X_k\}_{k \in S}$. We consider the scenario where a latent observation model parameter $\beta_S$ may exist, with corresponding $P(\beta_S)$ and $P(Y | X_S, \beta_S)$. Finally, let $\om$ index all sets of size $K$ among $N$ variables, such that $\om \in \left\{ 1,\ldots,\binom{N}{K} \right\}$ and the corresponding set is $S_\om$. In addition to conditional independence, the only other assumption we make is that $\om$ is chosen uniformly at random among $\binom{N}{K}$ sets.

A simple example of the type of problems we consider is the CS model \cite{donoho}, where the observations are given by $Y = \langle X, \beta \rangle + W$ for a $K$-sparse vector $\beta$ with support $S$, support coefficients $\beta_S$ and noise $W$. Another example is the group testing model \cite{group_testing}, where $X$ is a Boolean test inclusion vector that determines whether an item is included in the test or not and $S$ is the set of defective items. The group testing model assumes that the test outcomes $Y$ are only dependent on the inclusion of defective items, given by $X_S$. Further examples and details can be found in \cite{arxiv}.

We observe a sequence of $T$ variable-observation pairs $(X^T, Y^T) = (X^{(t)}, Y^{(t)})_{t = 1:T}$, where we used $a\!:\!b$ to denote the sequence of integers $(k \in \mathbb{N}: a \leq k \leq b)$. A decoder $g(X^T,Y^T)$ outputs an estimate $\omh$ of index $\om$ and we aim to characterize the error probability $P_e$ that $\omh \neq \om$ to obtain conditions on $T$ for successful recovery, in terms of $K$, $N$ and other problem parameters.

We now present the bounds on sample complexity from \cite{arxiv}. The following result is a lower bound on the number of samples for recovery, which is the nonadaptive analogue of our main result, Theorem \ref{thm:aLB}.

\begin{theorem}[Nonadaptive lower bound]
  \label{thm:LB}
  Let $X^T = (X^{(1)},\ldots,X^{(T)})$ be generated IID across $t = 1:T$ according to $p(X)$. 
  Then, a lower bound (or a necessary condition) on the number of samples required for $P_e$ to be asymptotically bounded away from zero is given by
  \begin{equation}
    T \geq \max_{\St \subset \So} \frac{\log\binom{N-|\St|}{K-|\St|}}{\Ist},
  \label{eq:LB} 
  \end{equation}
  where $\St$ is a proper subset of $\So$ and we define 
    \[ \Ist = I(X_{\So \setminus \St};Y | X_\St,\bS). \]
\end{theorem}

An upper bound on the number of samples is also presented in \cite{arxiv}. 
For the upper bound, it is further assumed that the variables are generated IID across both samples $t=1:T$ and variables $n=1:N$. The error probability of a Maximum Likelihood decoder is analyzed to obtain a sufficient condition for recovery. 
The lower bound given in Theorem \ref{thm:LB} is \emph{order-wise} tight as it matches the upper bound, when restricted to IID probability distributions on $X^T$, for $K$ not scaling with $N$ and provided that a mild condition on the mutual information is satisfied.

\section{A Lower Bound for Adaptive Recovery}

In this section we analyze the adaptive scenario where at each $t=1:T$, $X^{(t)}$ is given by a (possibly random) function $X^{(t)} = f_t(X^{(1:t-1)}, Y^{(1:t-1)})$. 
We state a lower bound on the number of samples in the adaptive case that holds for any distribution of $X^{(1)}$ and functions $f_t$. The bound depends on the distributions $p_t(X^{(t)})$ marginalized with respect to other sequence indices.

\begin{theorem}[Adaptive lower bound]
  \label{thm:aLB}
  Let $X^T = (X^{(1)},\ldots,X^{(T)})$ be generated such that each $X^{(t)}$ is a (random) function of $X^{(1:t-1)}$ and $Y^{(1:t-1)}$. 
  Then, a lower bound on the number of samples required for $P_e$ to be asymptotically bounded away from zero is given by
  \begin{equation}
    T \geq \max_{\St \subset \So} \frac{\log\binom{N-|\St|}{K-|\St|}}{\Isto}, 
  \label{eq:aLB} 
  \end{equation}
  where $\St$ is a proper subset and we define 
    \[ \Isto = \frac{1}{T} \sum_{t=1}^{T} I(X_{\So \setminus \St}^{(t)}; Y^{(t)} | X_\St^{(t)},\bS), \]
  as the average mutual information over the sequence $t=1:T$.
  Each term in the maximization above is also a lower bound.
\end{theorem}

In the nonadaptive case, $X^T$ is generated IID across samples $t=1:T$, therefore $\Isto = \Ist$ and the above bound reduces to the bound given in Theorem \ref{thm:LB}. This also holds for the adaptive case if the variables are chosen such that $X_\So^{(t)}$ is identically distributed across the sequence $t = 1:T$.

In order to obtain bounds for specific applications in the following section, we use two simple methods: We upper bound $\Ist$ directly for any $p(X)$ in group testing and 1-bit CS, which leads to a trivial upper bound on $\Isto$. For linear CS, we bound $\max_{p_t(X)} I(X_{\So \setminus \St}^{(t)}; Y^{(t)} | X_\St^{(t)},\bS)$ individually for $t=1:T$, which then gives an upper bound on $\Isto$.

\begin{IEEEproof}
    Let $\om \in \left\{1,2,\ldots, \binom{N}{K} \right\}$ be the true index of the salient set such that the condition $P(Y | X) = P(Y | X_\So)$ is satisfied. 
  Suppose a subset of the true support is revealed, denoted by $\St \subset \So$, so that only elements of the set ${\So \setminus \St}$ are left to be identified. Define $E$ as the binary error event that the estimate $\omh = g(X^T, Y^T)$ is not equal to $\om$ for a decoder $g$ and let $P_e$ be the probability of this event. 
  We start by writing the uncertainty of $\om$ given $\St$,
  \begin{align}
    H(\om | \St) & = H({\So \setminus \St}) = \log \binom{N-|\St|}{K-|\St|} \label{eq:ad1} \\
    & = H(\om | \omh, \St) + I(\om ; \omh | \St) \label{eq:ad2} \\
    & \leq H(\om | \omh, \St) + I(\om ; X^T, Y^T | \St) \label{eq:ad3},
  \end{align}
  where \eqref{eq:ad1} is due to $\om$ being chosen uniformly at random, \eqref{eq:ad2} follows from standard entropic identities and \eqref{eq:ad3} is due to data processing inequality and $\omh$ being a function of $X^T$ and $Y^T$. 
  We analyze the first term and write the following inequality:
  \begin{align}
    H(\om | \omh, \St) & = H(E, \om | \omh, \St) - H(E | \om, \omh, \St) \label{eq:ad4} \\
    & = H(E | \omh, \St) + H(\om | E, \omh, \St) \label{eq:ad5} \\
    & \leq 1 + (1 - P_e) \, 0 + P_e \, H(\om | \St) \label{eq:ad6} \\
    & = 1 + P_e \log\binom{N-|\St|}{K-|\St|}. \label{eq:ad7}
  \end{align}
  The two equalities \eqref{eq:ad4} and \eqref{eq:ad5} again follow from standard identities by noting that the second term in \eqref{eq:ad4} is zero since $E$ is determined completely by $\om$ and $\omh$. \eqref{eq:ad6} follows by upper bounding the entropy of a binary variable by 1, expanding the conditional entropy for $E = 0$ and $E = 1$, noting that $E = 0$ implies $\om = \omh$ and removing the conditioning on $\omh$ on the last term. 
  
  We now look at the second term, $I(\om ; X^T, Y^T | \St)$. We first note the following:
  \begin{align*}
    I(\om ; X^T, Y^T | \St) & \leq I(\om ; X^T, Y^T, \bS | \St) \\
    & = I(\om ; X^T, Y^T | \St, \bS) + I(\om ; \bS | \St) \\
    & = I(\om ; X^T, Y^T | \St, \bS),
  \end{align*}
  where the last equality follows from the independence of $\om$ and $\bS$. For this term, we can then write,
  \begin{align}
  & I(\om ; X^T, Y^T | \St, \bS) = H(X^T, Y^T | \St, \bS) \nonumber \\
    & \quad - H(X^T, Y^T | \om, \bS) \label{eq:ad8} \\
    & = \sum_{t=1}^T H(X^{(t)}, Y^{(t)} | X^{(1:t-1)}, Y^{(1:t-1)}, \St, \bS) \nonumber \\
    & \qquad - H(X^{(t)}, Y^{(t)} | X^{(1:t-1)}, Y^{(1:t-1)}, \om, \bS) \label{eq:ad9} \\
    & = \sum_{t=1}^T \left( H(Y^{(t)} | X^{(t)}, \St, \bS) \nonumber \right. \\
    & \qquad + \left. H(X^{(t)} | X^{(1:t-1)}, Y^{(1:t-1)}, \St, \bS) \right) \nonumber \\
    & \qquad - \left( H(Y^{(t)} | X^{(t)}, \om, \bS) \nonumber \right. \\ 
    & \qquad + \left. H(X^{(t)} | X^{(1:t-1)}, Y^{(1:t-1)}, \om, \bS) \right) \label{eq:ad10} \\
    & \leq \sum_{t=1}^T I(X^{(t)}_{\So \setminus \St} ; Y^{(t)} | X^{(t)}_\St, \bS) \nonumber \\ 
    & \quad + I(X^{(t)} ; {\So \setminus \St} | X^{(1:t-1)}, Y^{(1:t-1)}, \St, \bS) \label{eq:ad12} \\
    & = \sum_{t=1}^T I(X^{(t)}_{\So \setminus \St} ; Y^{(t)} | X^{(t)}_\St, \bS) \triangleq T \Isto, \label{eq:ad13} 
  \end{align}
  where \eqref{eq:ad9} follows from the chain rule of entropy and \eqref{eq:ad10} follows from the fact that $Y^{(t)}$ is independent of $X^{(t^\prime)}$ and $Y^{(t^\prime)}$ for $t^\prime \neq t$ given $X^{(t)}$ and $\bS$. We note that $X^{(t)}_\St$ is a function of $X^{(t)}$ and $\St$, and $Y^{(t)}$ depends only on $X^{(t)}_\om$ given $X^{(t)}$ and $\om$ to obtain \eqref{eq:ad12}. Finally, as $X^{(t)}$ only depends on $(X^{(1:t-1)}, Y^{(1:t-1)})$, the second term is zero and \eqref{eq:ad13} follows. 
  
  Putting together \eqref{eq:ad3}, \eqref{eq:ad7} and \eqref{eq:ad13}, we can write
    \[ \log\binom{N-|\St|}{K-|\St|} \leq 1 + P_e \log\binom{N-|\St|}{K-|\St|} + T \Isto, \]
  which leads to
    \[ P_e \geq 1 - \left( \frac{T \Isto + 1}{\log\binom{N-|\St|}{K-|\St|}} \right),   \]
  and thus for $P_e$ not to be strictly positive, we need
    \[ T \geq \frac{\log\binom{N-|\St|}{K-|\St|} - 1}{\Isto}.  \]  
  
  Considering all proper subsets $\St \subset \So$, 
  a lower bound on $T$ for recovery is then given by \eqref{eq:aLB}.
\end{IEEEproof}

While the proof is inspired by the feedback proof of \cite{coverbook}, it is fundamentally different since we consider extra latent parameters $\bS$, we have the extra overlap terms $X_\St$ and we explicitly assume variables depend not only on past outputs but also on past inputs.

\section{Applications}

In this section we discuss the implications of the adaptive lower bound for some applications. We will first present results for group testing and 1-bit CS; then we will look at the linear CS model.

\subsection{Group Testing}

Group testing is the problem of identifying a set of ``defective'' items from a larger set, where group tests can be performed which results in a positive outcome if and only if a defective item is included in the test group. Formally, we let $X^T$ denote the Boolean test inclusion matrix for $N$ items and $T$ tests and $Y^T$ denote the binary outcomes of $T$ tests, where each test outcome is given by the formula 
\[ Y^{(t)} = \bigvee_{k \in S} X_k^{(t)}. \]

\begin{theorem}
  $T = \Theta(K \log(N/K))$ is a lower bound on the number of tests to recover the defective set $S$ with an arbitrarily small error probability using adaptive testing.
\end{theorem}

\begin{IEEEproof}
  Consider the case $\St = \emptyset$. Then note that $\Isto \leq \frac{1}{T} \sum_{t=1}^T H(Y^{(t)}) \leq 1$ as $Y$ is binary. Since $\log\binom{N}{K} = \Theta(K \log(N/K))$, it follows from Theorem \ref{thm:aLB} that $T = \Theta(K \log(N/K))$ is a lower bound. 
\end{IEEEproof}

The same lower bound was shown for nonadaptive group testing in by the authors in \cite{group_testing,arxiv} and for adaptive in \cite{aldridge,aldridge2}\footnote{\cite{aldridge} inadvertently makes a strong assumption that the variables are identically distributed over the sequence, which is not necessarily true for adaptive testing and leads to the identity $\Isto = \Ist$. In subsequent work \cite{aldridge2} the authors obtain the same lower bound through a different argument.}. 
Asymptotically matching upper bounds have also been shown for nonadaptive and adaptive testing, see \cite{aldridge2}. In fact, the lower bound can be achieved by choosing entries of $X^T$ IID $\sim$ Bernoulli$(1/K)$ for $K = o(N)$ \cite{group_testing}. Therefore we observe that adaptivity cannot improve performance asymptotically in this sparse recovery problem.

Note that versions of the group testing problem with noisy test outcomes can also be considered as in \cite{group_testing} and our results for adaptive testing can be extended to these models.

\subsection{1-bit Compressive Sensing}

1-bit CS \cite{gupta} is interesting as the extreme case of quantized CS models which are of practical importance in many real world applications. Mathematically, we have 
\[ Y^T = Q(X^T \beta + W^T), \]
where $X^T$ is a $T \times N$ sensing matrix with $t$-th row corresponding to $X^{(t)}$, $\beta$ is a $K$-sparse $N \times 1$ vector with support $S$ and $W^T$ is an IID noise vector. $Q(\cdot)$ is a 1-bit quantizer which outputs $1$ if the input is nonnegative and $0$ otherwise, for each element in the input vector. 

\begin{theorem}
$T = \Theta(K \log(N/K))$ is a lower bound on the number of measurements to recover the support $S$ of $\beta$ using adaptive measurements with an arbitrarily small error probability.
\end{theorem}

The proof is the same as the proof for group testing since $Y^{(t)}$ are binary measurements. Matching upper bounds for noiseless and noisy variants of the problem (with sufficiently high SNR) have been shown  by the authors in \cite{gupta} and \cite{arxiv} using IID Gaussian measurement matrices. Therefore we have shown that support recovery performance in 1-bit CS cannot be increased asymptotically using adaptive measurements in those SNR regimes.

\subsection{Compressive Sensing}

We now look at the CS problem with measurement noise. We have the normalized model \cite{shuchin,arxiv},
\[ Y^T = X^T \beta + W^T, \]
where $X^T$ is the $T \times N$ sensing matrix, $\beta$ is a $K$-sparse vector of length $N$ with support $S$ and $Y^T$ is the observation vector of length $T$. $W^T$ is an IID Gaussian noise vector with variance $1/\SNR$. We assume w.l.o.g.\ (since we are obtaining a lower bound) that $\beta_k \in \{-1, +1\}$ with equal probability and are IID for $k \in S$. We constrain the total power of the entries of $X^T$ similar to \cite{malloy,haupt}, where we assume $\sum_{n=1}^N E\left[ X_n^{(t)2} \right] = N P_t$ and $\sum_{t=1}^T P_t = 1$, to be consistent with \cite{arxiv} and \cite{shuchin}. As a special case, row-wise power constraints can be enforced by setting $P_t = \frac{1}{T}$. 
Note that this normalized model uses a different convention from the fixed noise variance models used in \cite{malloy,haupt,davenport} where bounds on the minimum magnitude of support coefficients are derived instead of $\SNR$.

In order to obtain a valid lower bound for all distributions on $X$, we optimize $\IsttS$ over probability distributions $p_t(X^{(t)})$ satisfying the power constraint, simultaneously for all $\St \subset S$. 
Let $\Sp = S \setminus \St$ and note that
\begin{align*}
  \IsttS &  \\
  & \hspace{-60pt} = h(Y^{(t)} | X_\St^{(t)}, \beta_S) - h(Y^{(t)} | X_S^{(t)}, \beta_S) \\
  & \hspace{-60pt} = h(X_S^{(t)\top} \beta_S + W^{(t)} | X_\St^{(t)}, \beta_S) \\
  &\hspace{-48pt} - h(X_S^{(t)\top} \beta_S + W^{(t)} | X_S^{(t)}, \beta_S) \\
  & \hspace{-60pt} = h(X_{\Sp}^{(t)\top} \beta_{\Sp} + W^{(t)} | X_\St^{(t)}, \beta_{\Sp}) - h(W^{(t)}).
\end{align*}

\begin{lemma}
  $p_t(X^{(t)})$ that maximizes $\min_{\St} \IsttS$ subject to power constraints sets $\{X_k^{(t)}\}_{k \not\in S}= 0$ and is jointly Gaussian in $X_S^{(t)}$, with zero mean and covariance matrix $\Sigma^{(t)}$ with diagonals $\frac{N P_t}{K}$ and the off-diagonals equal to some scalar $\rho$.
\end{lemma}

We do not prove the lemma due to space constraints, however it follows from the fact that $\beta_S$ is IID and symmetric around zero, the maximization is over all subsets $\St \subset S$ and since an entropy is being maximized subject to power constraints.

Then, removing the conditioning on $X_\St$, $\IsttS$ can be upper bounded by
\begin{align*}
  \IsttS & \\
  & \hspace{-60pt} \leq \frac{1}{2}E_{\beta_{\Sp}} \left[ \ln\left( 2 \pi e \left[\beta_{\Sp}^\top \Sigma_\Sp^{(t)} \beta_{\Sp} + \frac{1}{\SNR} \right]\right)\right] \\
  & \hspace{-48pt} - \frac{1}{2} \ln\left( 2 \pi e \frac{1}{\SNR} \right) \\
  & \hspace{-60pt} = \frac{1}{2}E_{\beta_{\Sp}} \left[ \ln\left( 1 + \SNR \beta_{\Sp}^\top \Sigma_\Sp^{(t)} \beta_{\Sp} \right)\right] \\
  & \hspace{-60pt} \leq \frac{1}{2} \ln \left( 1 + \SNR E\left[ \beta_{\Sp}^\top \Sigma_\Sp^{(t)} \beta_{\Sp} \right] \right),
\end{align*}
due to Jensen's inequality, where $\Sigma_\Sp$ is the submatrix of $\Sigma$ corresponding to the indices in $\Sp$. Since $\Sigma_\Sp$ is a circulant matrix, we can write $\Sigma_\Sp = F \Lambda F^\star$ where $F$ is the unitary DFT matrix and $F^\star$ is its conjugate transpose. $\Lambda$ is a diagonal matrix with its first element equal to $\frac{N P_t}{K} + (|\Sp|-1)\rho$ and other $|\Sp| - 1$ diagonals equal to $\frac{N P_t}{K}-\rho$. It is also easy to show that $\widetilde{\beta_\Sp} = F^\star \beta_\Sp$ is also IID and has variances equal to $1$. Then it follows that, independent of the value of $\rho$, we have
 \[ \IsttS \leq \frac{1}{2} \ln\left( 1 + \SNR \frac{(K-|\St|) N P_t}{K} \right). \]

For $\Isto$ we can then write
\begin{align*}
  \Isto & \leq \frac{1}{T} \sum_{t=1}^T \frac{1}{2}\ln\left( 1 + \SNR \frac{(K-|\St|) N P_t}{K} \right) \\
  & \leq \frac{1}{2} \ln\left( 1 + \SNR \frac{(K-|\St|) N}{K T} \right), 
\end{align*}
where the second inequality follows from the fact that the sum is maximized by $P_t = \frac{1}{T}$ for all $t$ subject to the constraint $\sum_t P_t = 1$. Note that this implies distributing equal power to all $T$ rows. Therefore the same bound holds for the case where row-wise power is constrained instead of the total power of all entries of $X^T$.

We then evaluate \eqref{eq:aLB} with the above bound on $\Isto$ for $(T, \SNR)$ pairs to obtain the following theorem. Note that we have reduced the maximization over $\St \subset S$ in \eqref{eq:aLB} to a maximization over $i = K - |\St|$. 

\begin{theorem}\label{thm:CS}
  A necessary condition on a $(T, \SNR)$ pair for exact support recovery of $\beta$ with adaptive measurements and an arbitrarily small error probability is $T \log\left(1 + \SNR \frac{iN}{KT} \right) = \Omega(i \log(N/i))$ for all $i \in \{1,\ldots,K\}$.
\end{theorem}

Below corollary then follows by noting that the left-hand side is increasing in $T$ and letting $T$ grow to infinity independent of other quantities. Note that the theorem states that the condition should hold for all $i$ and we obtain the following bound for the case $i=1$. 

\begin{corollary}\label{thm:SNR}
  $\SNR = \Omega\left(\frac{K \log N}{N}\right)$ is a necessary condition for exact support recovery with adaptive measurements and an arbitrarily small error probability.
\end{corollary}

Our result is unique since the lower bound on $\SNR$ and the relationship between different $T$ and $\SNR$ can be explicitly characterized. Our result proves that for the linear sparsity regime $K = \Theta(N)$, $\SNR = \Theta(\log N)$ is necessary and $T = \Theta(N)$ is necessary for that $\SNR$. This shows that adaptivity does not help compared to nonadaptive since $\SNR = \Theta(\log N)$ is necessary and $T = \Theta(N)$ is achievable in that case \cite{shuchin}.

Corollary \ref{thm:SNR} implies that the necessary condition on $\SNR$ for adaptive measurements is possibly more relaxed for $K = o(N)$, which is also implied by the best known adaptive lower bound $\Omega(\log K)$ \cite{malloy}. While our bound is weaker in the sparser regimes, it is possible that a tighter analysis of the mutual information may lead to a comparable bound. Similarly, Theorem \ref{thm:CS} implies that it might be possible to recover $S$ with fewer measurements with the same noise levels for $K=o(N)$: An example is $T = \Theta(K)$ and $\SNR = \Omega(\log N)$, compared to $T = \Theta(K \log(N/K))$ for nonadaptive recovery; or $T = \Theta(K \log N)$ for $\SNR = \Omega(\log K)$, which is achievable for $K=o(N)$ \cite{malloy}. Therefore, in contrast to the group testing and 1-bit CS examples, in linear CS we have shown that there might be room for improvement using adaptive measurements in the sublinear sparsity regime. This result is consistent with previous lower bounds for adaptive CS \cite{haupt, castro}. 

Finally, note that performance improvements would not be possible if we constrained the power of each element of the sensing matrix individually or constrained the total power of $\langle X, \beta \rangle$, as in both cases it would not be possible to improve measurements by ``concentrating'' power on $X_S$ with the information from previous measurements.

\section{Discussion}

Considering Theorem \ref{thm:aLB} and the applications discussed above, we can argue that sample complexity is minimized in cases where the average information in the sequence $\Isto$ can approach the maximum value of $\Ist$ maximized over $p(X)$. However, in models where the only difficulty is the uncertainty in the observation model $P(Y | X_\So)$ and there are no total-power restrictions on measurements $X$ (even if there are element-wise restrictions), an optimal $p(X)$ can be determined beforehand. This $p(X)$ would maximize $\Ist$ for all possible sets $\om$ simultaneously, eliminating the need for adaptive measurements. Group testing is an example of such a problem where there are no restrictions on the Boolean testing matrix and the only uncertainty is due to the observation model, and hence we see no gains from adaptivity asymptotically.

However, for problems with a total-power constraint (such as the linear CS model), adaptivity can help by obtaining more information on $\om$ and transferring power from less likely candidates for $\So$ to more likely ones, therefore increasing the ``effective SNR'' as the measurement sequence progresses. Since nonadaptive measurements do not have any prior information on $\om$, they attempt to distribute power evenly over all $N$ candidate indices so they cannot achieve the same performance asymptotically, at least for sublinear sparsity.

1-bit CS is an interesting example since $\SNR$ and the total power of measurement vectors are still important but we observe that for sufficiently high $\SNR$ the binary output constrains performance more than the measurement power. Therefore nonadaptive methods can achieve the adaptive lower bound asymptotically. However, it is an open question whether adaptivity can increase asymptotic performance for lower $\SNR$ values similar to linear CS.

\section*{Acknowledgements}

This work was supported by NSF Grant 0932114 and NSF Grant CCF-1320547.

\bibliographystyle{IEEEtran}
\bibliography{references_adaptive}

\end{document}